\documentclass{amsart}

\usepackage {epsfig}
\usepackage {graphicx}

 \newtheorem{thm}{Theorem}
 
 \newtheorem{lemma}[thm]{Lemma}
 
 \newtheorem{theorem}[thm]{Theorem}
 \theoremstyle{definition}
 
 \theoremstyle{remark}

 \newcommand{\Real}{\mathbb{R}}

\usepackage{graphicx}
\begin{document}

\title[Synchronization of class I neurons]
{Synchronization in model networks of class I neurons}

\author{Guy Katriel}

\address{}

\email{haggaik@wowmail.com}


\begin{abstract}
We study a modification of the canonical model for networks of
class I neurons derived by Hoppensteadt and Izhikevich, in which
the `pulse' emitted by a neuron is smooth rather than a
$\delta$-function. We prove two types of results about
synchronization and desynchronization of such networks, the first
type pertaining to `pulse' functions which are symmetric, and the
other type in the regime in which each neuron is connected to many
other neurons.

\end{abstract}

\maketitle

\section{introduction}

In the work of Hoppensteadt and Izhikevich  \cite{hibook,iz} a
model for networks of class I neurons was derived, which is
canonical in the sense that a huge class of detailed and
biologically plausible models of weakly-connected neurons near a
`saddle-node on limit-cycle' bifurcation can be reduced to it by a
piecewise-continuous change of variables. We refer to \cite{iz}
for the precise conditions under which this reduction is valid,
and to \cite{hibook}, chapter 8, for details of the derivation.
The canonical model is described by the equations
\begin{equation}
\label{pc}\theta_i'=h(r_i;\theta_i)+\sum_{j=1}^n
{w(s_{ij};\theta_i)\delta(\theta_j-\pi)},\;\;\;1\leq i\leq n,
\end{equation}
where $\theta_i$ is the phase of the $i$-th neuron, and in which
$$h(r;\alpha)=(1-\cos(\alpha))+(1+\cos(\alpha))r$$
describes the internal dynamics of the $i$-th neuron (the
Ermentrout-Kopell model), which is excitable if $r<0$ and
oscillatory if $r>0$, $w(s;\alpha)$, which describes the response
of neuron $i$ to a pulse from neuron $j$ is the $2\pi$-periodic
extension of the function defined for $\alpha\in [-\pi,\pi]$ by
\begin{equation}
\label{defw}
w(s;\alpha)=2\arctan\Big(\tan(\frac{\alpha}{2})+s\Big)-\alpha,
\end{equation}
where $s$ is a parameter such that the connection from neuron $j$
to neuron $i$ is excitatory if $s_{ij}>0$ and inhibitory if
$s_{ij}<0$ . An important property of $w$ is that
\begin{equation}
\label{van} w(s;\pi)=0 \;\;\;\forall s\in\Real
\end{equation}
which means that the neurons are unaffected by input from other
neurons at the moment they are firing. $\delta$ is the Dirac
delta-function, and the term $\delta(\theta_j-\pi)$ indicates that
the neurons fire whenever their phase is $\pi$. This $\delta$-term
is in fact an approximation of a smooth pulse-like function up to
order $O(\sqrt{\epsilon}\log(\epsilon))$, where $\epsilon$ is the
strength of connections in the `unreduced' model, which is assumed
to be small.

A central result of \cite{iz}, confirming and extending previous
numerical and analytical work \cite{hansel,ermentrout} is that
networks of class I neurons {\it{desynchronize}}.

In this work we analyze a modified model in which the
$\delta$-term is replaced by a smooth $2\pi$-periodic `pulse-like'
function $P(\beta)$ satisfying
\begin{equation}\label{pp}
P(\beta)>0\;\;\;\forall \beta\in\Real.
\end{equation}
We thus consider networks described by the equations
\begin{equation}
\label{mod}
\theta_i'=h(r_i;\theta_i)+\sum_{j=1}^{n}{w(s_{ij};\theta_i)P(\theta_j)},\;\;\;1\leq
i\leq n
\end{equation}

In fact the senses in which we shall require $P$ to be
`pulse-like' are very weak; indeed a central motivation for
investigating the modified model, besides the fact that the model
(\ref{mod}), for an appropriate choice of the function $P$, is
actually more precise than (\ref{pc}), is that studying
(\ref{mod}) allows us to examine the robustness - and limitations
- of the desynchronization results obtained before for the
pulse-coupled model. The use of a smooth `pulse-like' function $P$
also requires and allows us to use tools of smooth analysis,
leading to interesting mathematical investigations and results. We
shall also prove results in which the specific functional forms
$h(r;\alpha)$, $w(s;\alpha)$ are replaced by general functions
satisfying some of the key qualitative properties of these
particular functions. These more general results contribute to
elucidating what aspects of the model (\ref{pc}) are responsible
for its dynamical behavior.

As in \cite{iz}, we investigate the issue of synchronization: does
there exist a {\bf{synchronized oscillation}} of the system,
{\it{i.e.}}, a solution for which
$\theta_i(t)=\overline{\theta}(t)$ for all $1\leq i \leq n$, with
\begin{equation}
\label{osc} \overline{\theta}(t+T)=\overline{\theta}(t)+2\pi
\;\;\; \forall t\in \Real,
\end{equation}
for some $T>0$, and if it exists, is it {\it{stable}}? If the
answer to both these questions is positive, we say that the
network {\bf{synchronizes}}. Otherwise we say that the network
{\bf{desynchronizes}}.

After some preliminaries in sections \ref{ex},\ref{st}, we present
and prove our main results. In section \ref{de} we study the
behavior of the model that when the pulse function $P$ is
symmetric with respect to $\pi$ and increasing on $(0,\pi)$, and
show that when $r\geq 0$ we have desynchronization. In section
\ref{sdm} we deal with the case of large networks with many
connections per neuron, and prove that in this regime, when
$P'(\pi)\neq 0$ then synchronization/desynchronization depends
only on the local behavior of the `pulse-like' function $P$ near
$\pi$ - if $P'(\pi)>0$ then desynchronization occurs, while if
$P'(\pi)<0$ the synchronized oscillation is stable.

\section{existence of synchronized oscillations}
\label{ex}

In order to ensure existence of a synchronized oscillation, we
need to make some assumptions. As in \cite{iz}, we assume that the
internal dynamics of the oscillators are identical, that is,
\begin{equation}
\label{re} r_i=r,\;\;\;1\leq i\leq n
\end{equation} In \cite{iz} it
is assumed that $s_{ij}=s$ for all $i,j$. In fact less stringent
assumptions are sufficient to ensure existence of a synchronized
oscillation. We assume that for any $i,j$,
\begin{equation}
\label{as} s_{ij}=0\;or\;s.
\end{equation}
We note that
\begin{equation}
\label{zz}w(0;\alpha)=0\;\;\;\forall \alpha\in\Real
\end{equation}
which means that when $s_{ij}=0$ neuron $j$ does not influence
neuron $i$. Thus assumption (\ref{as}) allows for an arbitrary
connectivity structure in the network (which can be represented by
a directed graph with the neurons as nodes and an arrow from
neuron $j$ to neuron $i$ iff $s_{ij}=s$), stipulating only that
those connections which exist are identical. This means that we
can rewrite (\ref{mod}) in the form
\begin{equation}
\label{con}
\theta_i'=h(r;\theta_i)+\sum_{j=1}^{n}{c_{ij}w(s;\theta_i)P(\theta_j)},\;\;\;1\leq
i\leq n
\end{equation}
where $c_{ij}=0\;or\;1$ for all $i,j$.

In the case of the `delta-coupled' system (\ref{pc}), under
conditions (\ref{re}), (\ref{as}), the assumption $r>0$ is
necessary and sufficient to guarantee the existence of a
synchronized oscillation; indeed, due to (\ref{van}), the coupling
term vanishes identically if we set $\theta_i=\overline{\theta}$
($1\leq i\leq n$), so that the system (\ref{pc}) reduces to
$$\overline{\theta}'=(1-cos(\overline{\theta}))+(1+cos(\overline{\theta}))r,$$
which has a solution satisfying (\ref{osc}) iff $r>0$. In the case
of the `smooth' system (\ref{mod}), these assumptions are not
sufficient, and in order to guarantee that upon substituting
$\theta_i=\overline{\theta}$ we get the {\it{same}} equation for
$\overline{\theta}$ from each of the equations (\ref{con}) we need
to assume further that the number of neurons affecting each neuron
is the same, so that the network is {\bf{$k$-regular}} in the
sense that each node in the corresponding directed graph has $k$
incoming arrows, or in other terms that
\begin{equation}
\label{reg} \sum_{j=1}^n{c_{ij}}=k,\;\;\;1\leq i\leq n
\end{equation}
(note that, in particular, the `all-to-all' coupling case
considered in \cite{iz} satisfies this assumption). If we assume
(\ref{reg}) holds then upon substituting
$\theta_i=\overline{\theta}$ ($1\leq i\leq n$) in any of the
equations (\ref{con}) we obtain
$$\overline{\theta}'=h(r;\overline{\theta})+kw(s;\overline{\theta})P(\overline{\theta}).
$$
This equation will have a solution satisfying (\ref{osc}) iff
\begin{equation}
\label{apos} h(r;\theta)+kw(s;\theta)P(\theta)>0 \;\;\;\forall
\theta\in\Real
\end{equation}
(and this synchronized oscillation is unique up to
time-translations). Whether this condition holds depends on the
parameters $r,s,k$. Using the properties of $h(r;\alpha)$,
$w(s;\theta)$ and (\ref{pp}), we can easily identify some regimes
in which (\ref{apos}) holds or does not hold, so that we have the
following result about existence.

\begin{theorem}
\label{exist} Consider the network described by (\ref{con}), and
assume that it is $k$-regular.

\noindent (i) When $r\geq 0$ and $s>0$, a synchronized oscillation
exists.

\noindent (ii) When $r\leq 0$ and $s<0$, a synchronized
oscillation does {\it{not}} exist.

\noindent (iii) When $r>0$ and $s<0$:

(a) If $k$ is fixed then when $|s|$ is sufficiently small a
synchronized oscillation exists.

(b) If $s$ is fixed then for $k$ sufficiently large a synchronized
oscillation does {\it{not}} exist.

\noindent (iv) When $r<0$ and $s>0$:

(a) If $k$ is fixed and $|s|$ sufficiently small a synchronized
oscillation does {\it{not}} exist.

(b) If $s$ is fixed and $k$ is sufficiently large then a
synchronized oscillation exists.
\end{theorem}

\section{stability of the synchronized oscillation}
\label{st}

Stability of the synchronized oscillation - in the sense that its
Floquet multipliers have modulus less than $1$ - is a crucial
question: if the synchronized oscillation is unstable we shall
never observe synchronization. Since we are dealing here with
local stability, even when the synchronized oscillation is stable
we cannot guarantee that the network will synchronize whatever the
initial conditions, but we do know that it will synchronize for
some open set of initial conditions, in particular if these
initial conditions are sufficiently close to each other.

In order to investigate the stability of the synchronized
oscillation we shall use the following theorem, which is a
corollary of theorem 8 of \cite{katriel}, pertaining to networks
described by equations of the form
\begin{equation}
\label{conventional}
\theta_i'=h(\theta_i)+\sum_{j=1}^{n}{c_{ij}f(\theta_i,\theta_j)},\;\;\;1\leq
i\leq n.
\end{equation}
A network is said to be {\bf{irreducible}} if it is {\it{not
possible}} to partition the neurons into two disjoint sets
$S_1,S_2$ so that $c_{ij}=0$ for all $i\in S_1,\; j\in S_2$ (in
other words, in such a way that neurons in $S_2$ do not influence
neurons in $S_1$).

\begin{theorem}
\label{stability} Consider a network described by equations of the
form (\ref{conventional}), where $h$ is $2\pi$-periodic and
$f(\alpha,\beta)$ is $2\pi$-periodic in both variables. Assume
that $c_{ij}\geq 0$ for all $i,j$, that the network is irreducible
and satisfies (\ref{reg}) and that
\begin{equation}\label{key}
h(\theta)+kf(\theta,\theta)>0 \;\;\;\forall \theta\in \Real.
\end{equation}
Then a synchronized oscillation, unique up to time-translations,
exists and:

\noindent (i) The synchronized oscillation is stable if $\chi>0$,
where $\chi$ is defined by
\begin{equation}
\label{defchi} \chi=\int_0^{2\pi}{\frac{\partial f}{\partial
\beta}(\theta,\theta)
\frac{d\theta}{h(\theta)+kf(\theta,\theta)}}.
\end{equation}

\noindent (ii) The synchronized oscillation is unstable if
$\chi<0$.
\end{theorem}

\section{the case of a symmetric pulse}
\label{de}

In this section we make the following assumptions on the smooth
$2\pi$-periodic `pulse-like' function $P(\beta)$:
\begin{equation}\label{sym}
P(\pi+\theta)=P(\pi-\theta)\;\;\; \forall\theta\in\Real
\end{equation}
\begin{equation}\label{inc}
P'(\beta)>0\;\;\;\forall\alpha\in (0,\pi)
\end{equation}

Assumption (\ref{sym}) means that the pulse is symmetric with
respect to $\pi$, and we note that by periodicity it is equivalent
to
\begin{equation}\label{sym1}
P(\beta)=P(-\beta)\;\;\; \forall\theta\in\Real
\end{equation}
Under these assumptions we shall prove:
\begin{theorem}\label{sd}
If $r\geq 0$, $P$ satisfies (\ref{pp}),(\ref{sym}),(\ref{inc}) and
the network described by (\ref{con}) is regular and irreducible,
then:

\noindent (i) If $s>0$ the synchronized oscillation of (\ref{con})
exists and is unstable.

\noindent (ii) If $s<0$ then a synchronized oscillation of
(\ref{con}) may or may not exist, but if it exists, it is
unstable.
\end{theorem}

Theorem \ref{sd} is a consequence of the following more general
theorem pertaining to networks described by the equations
\begin{equation}
\label{cong}
\theta_i'=h(\theta_i)+\sum_{j=1}^{n}{c_{ij}w(\theta_i)P(\theta_j)},\;\;\;1\leq
i\leq n
\end{equation}
which does not assume the specific functional forms
$h(\alpha)=h(r;\alpha)$, $w(\alpha)=w(s;\alpha)$, but rather only
some of the qualitative properties of these functions.

\begin{theorem}
\label{gds} Assume the smooth $2\pi$-periodic functions $h,w,P$
satisfy (\ref{pp}), (\ref{sym}), (\ref{inc}) and
\begin{equation}\label{hp}
h(\alpha)>0\;\;\;\forall \alpha\in (0,\pi),
\end{equation}
\begin{equation}\label{hs}
h(\alpha)=h(-\alpha)\;\;\;\forall \alpha\in\Real,
\end{equation}
\begin{equation}\label{ww}
w(-\alpha)>w(\alpha)\;\;\;\forall \alpha\in (0,\pi),
\end{equation}
and that the network described by (\ref{cong}) is regular and
irreducible. Then if a synchronized oscillation of (\ref{cong})
exists, it is unstable.
\end{theorem}

All the assumptions of theorem \ref{gds} hold for the case
$h(\alpha)=h(r;\alpha)$, $w(\alpha)=w(s;\alpha)$, $r\geq 0$, so
that theorem \ref{sd} follows. The only assumption which is not
immediate to verify is (\ref{ww}). We need to show

\begin{lemma}
\label{wl} If $s\neq0$ then
\begin{equation}\label{ww1}
w(s;-\alpha)>w(s;\alpha)\;\;\;\forall \alpha\in (0,\pi),
\end{equation}
\end{lemma}

To prove lemma \ref{wl}, we note that using the definition of
$w(s;\alpha)$ given by (\ref{defw}), (\ref{ww1}) is equivalent to
the inequality
$$
\arctan(s+\tan(\frac{\alpha}{2}))-\arctan(s-\tan(\frac{\alpha}{2}))<\alpha\;\;\;\forall
s\neq 0,\; \alpha\in(0,\pi).
$$
Setting $u=\tan(\frac{\alpha}{2})$, this inequality is equivalent
to
\begin{equation}\label{inw}
\arctan(s+u)-\arctan(s-u)<2\arctan(u)\;\;\;\forall s\neq 0,\; u>0.
\end{equation}
To prove (\ref{inw}), we {\it{fix}} an arbitrary $u>0$ and we
define
$$g(s)=\arctan(s-u)-\arctan(s+u)+2\arctan(u).$$
We need to show that $g(s)>0$ for all $s\neq 0$. We note that
$g(0)=0$ so that it suffices to show that $g'(s)>0$ for $s>0$ and
$g'(s)<0$ for $s<0$. But
$$g'(s)=\frac{1}{1+(s-u)^2}-\frac{1}{1+(s+u)^2},$$
and it is easy to check that the expression on the right-hand side
is positive when $u>0$, $s>0$ and negative when $u>0$, $s<0$,
completing the proof of lemma \ref{wl}.

\vspace{0.5cm} We now prove theorem \ref{gds}. Assume that a
synchronized oscillation of (\ref{cong}) exists, which is
equivalent to the assumption
\begin{equation}\label{es}
h(\theta)+kw(\theta)P(\theta)>0\;\;\;\forall \theta\in\Real.
\end{equation}
By theorem \ref{stability}, with
$f(\alpha,\beta)=w(\alpha)P(\beta)$, in order to determine the
stability of the synchronized oscillation we need to determine the
sign of
$$\chi=\int_{-\pi}^{\pi}{\frac{w(\theta)P'(\theta)d\theta}{h(\theta)+
kw(\theta)P(\theta)}}.$$ We define
$$F(\theta)=\frac{w(\theta)}{h(\theta)+
kw(\theta)P(\theta)},$$ so that we can write
$$\chi=\int_{-\pi}^{\pi}{F(\theta)P'(\theta)d\theta}.$$
We claim that
\begin{equation}\label{fi}
F(-\theta)>F(\theta)\;\;\;\forall \theta\in (0,\pi).
\end{equation}
To show this, we fix an arbitrary $\theta\in (0,\pi)$, and define
the function
$$\Lambda(x)=\frac{x}{h(\theta)+kP(\theta)x},$$
it is easy to verify (in view of (\ref{hp})) that $\Lambda(x)$ is
an increasing function on the interval $(x_0,\infty)$, where
$$x_0=-\frac{h(\theta)}{kP(\theta)}.$$
By (\ref{hp}),(\ref{es}) we have $w(\theta)>x_0$, so (\ref{ww})
implies that $\Lambda(w(-\theta))>\Lambda(w(\theta))$, which, in
view of (\ref{sym1}) and (\ref{hs}), is equivalent to (\ref{fi}).
We also note that, from (\ref{sym1}) we have
$P'(-\theta)=-P'(\theta)$ for all $\theta$, hence
$$\int_{-\pi}^{0}{F(\theta)P'(\theta)d\theta}
=\int_{0}^{\pi}{F(-\theta)P'(-\theta)d\theta}=
-\int_{0}^{\pi}{F(-\theta)P'(\theta)d\theta}$$ so that
$$\chi=\int_{-\pi}^{0}{F(\theta)P'(\theta)d\theta}+
\int_{0}^{\pi}{F(\theta)P'(\theta)d\theta}=
\int_{0}^{\pi}{(F(\theta)-F(-\theta))P'(\theta)d\theta}.$$ Using
this, (\ref{inc}) and (\ref{fi}), we obtain $\chi<0$ which
concludes the proof of theorem \ref{gds}.

\vspace{0.5cm} In theorem \ref{sd} we assumed $r\geq 0$. We now
examine the case $r<0$. If also $s<0$ then by part (ii) of theorem
\ref{exist}, there does not exist a synchronized oscillation. We
thus assume $s>0$ so that a synchronized oscillation may exist,
for example by part (iv)(b) of theorem \ref{exist} it exists for
$k$ sufficiently large. In the case $r<0$, $s>0$, in contrast to
the result of theorem \ref{sd}, it is possible for the conditions
(\ref{pp}),(\ref{sym}),(\ref{inc}) to hold and for the
synchronized oscillation to be stable. An example is
$r=-\frac{1}{2}$, $s=1$, $P(\beta)=2-\cos(\beta)$, $k=1$. One can
check that (\ref{es}) holds by graphing the function on the
right-hand side, so that a synchronized oscillation exists.
Computing $\chi$ numerically one finds $\chi=0.085..>0$, so that
the synchronized oscillation is stable. Thus in this example, in
which the uncoupled oscillators are excitable and the coupling is
excitatory, the coupling induces synchronization - whereas by
theorem \ref{sd} (under the stated assumptions on $P$), when the
uncoupled oscillators are oscillatory an excitatory coupling
induces desynchronization. To add further surprise, if in the
above example we increase $k$ to $k=2$, a synchronized oscillation
still exists, but now $\chi<0$ so that it is unstable. Thus, in
this example, increasing the connectivity of the network
{\it{desynchronizes}} the network which was synchronized by
sparser coupling!

\vspace{0.5cm} Let us note, in concluding our investigation of the
case of a symmetric pulse, that the property (\ref{ww}) is a key
element in the proof of the desynchronization result. Indeed,
following the same argument, we can see that if the function $w$
satisfies the reverse inequality $w(-\alpha)<w(\alpha)$
($\alpha\in (0,\pi)$) then, under the same assumptions on $P$ and
$h$, we will have synchronization. In this connection, it is
interesting to examine the simplified version of the model
(\ref{pc}), as presented in \cite{iz}, which consists in replacing
$w(s;\alpha)$ by
\begin{equation}\label{simplified}
\overline{w}(s;\alpha)=s(1+cos(\alpha)).
\end{equation}
This is the first term in the Taylor expansion of $w(s;\alpha)$
with respect to $s$, so that it is approximately  valid for small
$s$. Since $\overline{w}(s;\alpha)=\overline{w}(s;-\alpha)$, it is
immediate that whenever the pulse is symmetric the integral
defining $\chi$ vanishes. This means that all the Floquet
multipliers of the synchronized oscillation are $1$ (see the
formulas in \cite{katriel}) so that the synchronized oscillation
is `neutrally stable'. Thus, for the simplified model we have a
`weak desynchronization' result, but to obtain the strict
desynchronization the full model is necessary.

\section{the case of many connections}
\label{sdm}

In this section we consider the network (\ref{con}), assuming that
it is regular and irreducible and that $k$, the number of inputs
to each neuron, is large (since $k\leq n$ this implies that we are
dealing with large networks). We note that from parts (ii) and
(iii)(b) of theorem \ref{exist}, if $s<0$ then when $k$ is large a
synchronized oscillation does not exist, so in the following we
assume that $s>0$. We shall prove

\begin{theorem}
\label{lk} Fixing $r$ and $s>0$, and assuming $P$ is a smooth
$2\pi$-periodic function satisfying (\ref{pp}) and
\begin{equation}
\label{nmax} P'(\pi)\neq 0,
\end{equation}
we have the following results for any irreducible $k$-regular
network described by (\ref{con}), when $k$ is sufficiently large:

\noindent (i) If $P'(\pi)>0$, a synchronized oscillation exists
and is unstable.

\noindent (ii) If $P'(\pi)<0$, a synchronized oscillation exists
and is stable.
\end{theorem}

The striking characteristic of theorem \ref{lk} is that it shows
that when $P'(\pi)\neq 0$ then in the large-$k$ regime only the
local behavior of $P$ at $\pi$ is responsible for determining
synchronization/desynchronization (of course the special
significance of the value $\pi$ is that $w$ vanishes there).

Surprisingly, when $P'(\pi)=0$ (in particular when $P$ has a
maximum at $\pi$), the stability criterion is of a quite different
nature, in that it depends on the `global' behavior of the $P$:

\begin{theorem}
\label{lk1} Fix $r$ and $s>0$, and assume $P$ satisfies (\ref{pp})
and
\begin{equation}\label{pnd}
P'(\pi)=0.
\end{equation}
Define the quantity $\mu$ by the improper integral
\begin{equation}\label{defmus}
\mu=\int_{-\pi}^{\pi}{\frac{h(r;\theta)}{w(s;\theta)}\Big(\frac{1}{P(\theta)
}\Big)'d\theta} =\lim_{\gamma\rightarrow 0+}
{\int_{-\pi+\gamma}^{\pi-\gamma}
{\frac{h(r;\theta)}{w(s;\theta)}\Big(\frac{1}{P(\theta)
}\Big)'d\theta}} .\end{equation} Then for any irreducible
$k$-regular network described by (\ref{con}), when $k$ is
sufficiently large:

\noindent (i) If $\mu<0$, a synchronized oscillation exists and is
unstable.

\noindent (ii) If $\mu>0$, a synchronized oscillation exists and
is stable.
\end{theorem}

We note that the integrand in (\ref{defmus}) is singular at
$\pm\pi$, hence we have to define the integral as an improper
integral. It will follow from the proof of the theorem that $\mu$
is finite.

Theorems \ref{lk},\ref{lk1} follow from more general results about
networks of the form (\ref{cong}) where we do not make use of the
specific functional forms of $h(r,\alpha)$, $w(s,\alpha)$ in
(\ref{con}), but only need to assume some of their qualitative
properties. We note in particular that the simplified model
(\ref{simplified}) satisfies the assumptions of the theorems
below, so that the theorems \ref{lk},\ref{lk1} remain valid if
$w(s;\alpha)$ is replaced by (\ref{simplified}).

Theorem \ref{lk} follows from
\begin{theorem}
\label{nm} Let $h$ be a smooth $2\pi$-periodic function satisfying
\begin{equation}\label{hpp}
h(\pi)>0.
\end{equation}
Let $w$ be a smooth $2\pi$-periodic function satisfying
\begin{equation}\label{wn}
w(\alpha)> 0\;\;\;\forall \alpha\in(-\pi,\pi),
\end{equation}
\begin{equation}
\label{wzero} w(\pi)=0,\;\;\; w'(\pi)=0,\;\;\; w''(\pi)> 0
\end{equation}
Let $P(\beta)$ be a smooth $2\pi$-periodic function satisfying
(\ref{pp}) and (\ref{nmax}). Assume that the network described by
(\ref{cong}) is $k$-regular and irreducible. Then, for
sufficiently large $k$:

\noindent (i) If $P'(\pi)>0$ then the synchronized oscillation is
unstable.

\noindent (ii) If $P'(\pi)<0$ then the synchronized oscillation is
stable.

\end{theorem}

Theorem \ref{lk1} follows from
\begin{theorem}
\label{gent} Let $h$,$w$,$P$ be smooth $2\pi$-periodic functions
satisfying  (\ref{pp}), (\ref{pnd}), (\ref{hpp}), (\ref{wn}) and
(\ref{wzero}).

 Then, for sufficiently large $k$, any
$k$-regular irreducible network described by (\ref{cong}) has a
synchronized oscillation which is stable if $\mu>0$ and unstable
if $\mu<0$, where $\mu$ is defined by
\begin{equation}
\label{defmu} \mu=\int_{-\pi}^{\pi}
{\frac{h(\theta)}{w(\theta)}\Big(\frac{1}{P(\theta)
}\Big)'d\theta}=\lim_{\gamma\rightarrow 0+}
{\int_{-\pi+\gamma}^{\pi-\gamma}{
\frac{h(\theta)}{w(\theta)}\Big(\frac{1}{P(\theta)}\Big)'d\theta}}.
\end{equation}
\end{theorem}

\vspace{0.5cm}

To prove the above theorems, we need, according to theorem
\ref{stability}, to study the sign of
\begin{equation}
\chi=\chi(k)=\int_0^{2\pi}{\frac{w(\theta)P'(\theta)d\theta}{h(\theta)+
kw(\theta)P(\theta)}}
\end{equation}
As $k\rightarrow \infty$.

We note at the outset that we may assume, without loss of
generality, that we have
\begin{equation}
\label{hpos} h(\alpha)>0 \;\;\;\forall \alpha\in \Real.
\end{equation}
The reason for this is that if (\ref{hpos}) does not hold, we can
choose $k_0$ sufficiently large so that
$$\overline{h}(\alpha)\doteq
h(\alpha)+k_0w(\alpha)P(\alpha)>0\;\;\; \forall \alpha\in\Real$$
(this follows from the assumptions
(\ref{pp}),(\ref{hpp}),(\ref{wn})). Thus defining
$\overline{k}=k-k_0$ we have
$$\chi=\int_0^{2\pi}{\frac{w(\theta)P'(\theta)d\theta}{\overline{h}(\theta)+
\overline{k}w(\theta)P(\theta)}}.$$ Thus, we shall henceforth
assume (\ref{hpos}).

Setting $\epsilon=\frac{1}{k}$ we can write
$$\chi(\frac{1}{\epsilon})=\epsilon I(\epsilon)$$
where
$$I(\epsilon)=\int_0^{2\pi}{\frac{w(\theta)P'(\theta)}{h(\theta)}
\frac{d\theta}{\epsilon+\frac{w(\theta)P(\theta)}{h(\theta)}}}.$$
We need to determine the sign of $I(\epsilon)$ for $\epsilon>0$
small. We note first that, using periodicity
$$I(0)=\int_0^{2\pi}{\frac{P'(\theta)}{P(\theta)}d\theta}=
\int_0^{2\pi}{\frac{d}{d\theta}[\log(P(\theta))]}d\theta=0.$$
Thus, we need a delicate investigation in order to determine the
sign of $I(\epsilon)$ for small $\epsilon>0$. To simplify our
notation, we set
\begin{equation}\label{defa}
A(\theta)=\frac{w(\theta)P'(\theta)}{h(\theta)},
\end{equation}
\begin{equation}\label{defb}
B(\theta)=\frac{w(\theta)P(\theta)}{h(\theta)},
\end{equation}
so that
\begin{equation}\label{defj}
I(\epsilon)=\int_0^{2\pi}{\frac{A(\theta)d\theta}{\epsilon+B(\theta)}}.
\end{equation}

We now consider the case of theorem \ref{nm}, so that we assume
(\ref{nmax}) holds. Together with (\ref{wn}), (\ref{wzero}) and
(\ref{hpos}) this implies
$$A(\pi)=A'(\pi)=0,\;\;\; sign(A''(\pi))=sign(P'(\pi)),$$
\begin{equation}\label{bp}
B(\theta)>0\;\;\;\forall \theta\in(-\pi,\pi),
\end{equation}
\begin{equation}\label{bc}
B(\pi)=B'(\pi)=0,\;\;\; B''(\pi)>0.
\end{equation}
We may therefore write, for $\theta\in [0,2\pi]$:
$$A(\theta)=a(\theta-\pi)(\theta-\pi)^2,$$
$$B(\theta)=b(\theta-\pi)(\theta-\pi)^2,$$
where $a,b:[-\pi,\pi]\rightarrow \Real$ are continuous, and $b$ is
everywhere positive, and
\begin{equation}
\label{aa0} sign(a(0))=sign(P'(\pi)).
\end{equation}
We thus have
$$I(\epsilon)=\int_{-\pi}^{\pi}{\frac{a(\theta)\theta^2d\theta}
{\epsilon+b(\theta)\theta^2}}.$$ Decomposing into partial
fractions we have
$$I(\epsilon)=\int_{-\pi}^{\pi}{\frac{a(\theta)}{b(\theta)}d\theta}
-\epsilon\int_{-\pi}^{\pi}{\frac{a(\theta)}{b(\theta)}
\frac{d\theta}{\epsilon+b(\theta)\theta^2}}.$$ We have
$$\int_{-\pi}^{\pi}{\frac{a(\theta)}{b(\theta)}d\theta}=
\int_0^{2\pi}{\frac{A(\theta)}{B(\theta)}d\theta}=
\int_0^{2\pi}{\frac{P'(\theta)}{P(\theta)}d\theta}=0,$$ hence
$$I(\epsilon)=
-\epsilon\int_{-\pi}^{\pi}{\frac{a(\theta)}
{b(\theta)}\frac{d\theta}{\epsilon+b(\theta)\theta^2}}.$$
Substituting $\theta=\sqrt{\epsilon}u$ we obtain
$$I(\epsilon)=-\sqrt{\epsilon}
\int_{-\frac{\pi}{\sqrt{\epsilon}}}^{\frac{\pi}{\sqrt{\epsilon}}}
{\frac{a(\sqrt{\epsilon}u)}{b(\sqrt{\epsilon}u)}\frac{du}{1+b(\sqrt{\epsilon}u)u^2}},
$$
and by Lebesgue's dominated convergence theorem we have
$$\lim_{\epsilon\rightarrow
0+}{\frac{I(\epsilon)}{\sqrt{\epsilon}}}=-\frac{a(0)}{b(0)}\int_{-\infty}^{\infty}{\frac{du}{1+b(0)u^2}}.$$
Thus for small $\epsilon>0$ we have
$$sign(\chi)=sign(I(\epsilon))=-sign(a(0))=-sign(P'(\pi)),$$
and we obtain the result of theorem \ref{nm}. \vspace{0.5cm}

We turn to the proof of theorem \ref{gent}. We now have, from
(\ref{pp}),(\ref{pnd}),(\ref{wn}),(\ref{wzero}) and (\ref{hpos})
$$A(\pi)=A'(\pi)=A''(\pi)=0,$$
and also (\ref{bp}), (\ref{bc}). We may therefore write, for
$\theta\in [0,2\pi]$:
$$A(\theta)=a(\theta-\pi)(\theta-\pi)^3$$
$$B(\theta)=b(\theta-\pi)(\theta-\pi)^2$$
where $a,b:[-\pi,\pi]\rightarrow \Real$ are continuous, and $b$ is
everywhere positive. We can thus write $I(\epsilon)$ as
$$I(\epsilon)=\int_{-\pi}^{\pi}{\frac{a(\theta)\theta^3d\theta}{\epsilon+b(\theta)\theta^2}}.$$
We decompose into partial fractions,
obtaining
$$
I(\epsilon)=\int_{-\pi}^{\pi}{\frac{a(\theta)}{b(\theta)}\theta
d\theta}-\epsilon \int_{-\pi}^{\pi}{
\frac{a(\theta)}{b(\theta)}\frac{\theta
d\theta}{\epsilon+b(\theta)\theta^2}}.$$ We have
$$\int_{-\pi}^{\pi}{\frac{a(\theta)}{b(\theta)}\theta
d\theta}=\int_0^{2\pi}{\frac{A(\theta)}{B(\theta)}d\theta}=
\int_0^{2\pi}{\frac{P'(\theta)}{P(\theta)}d\theta}=0,$$ so that
$$
I(\epsilon)=-\epsilon \int_{-\pi}^{\pi}{
\frac{a(\theta)}{b(\theta)}\frac{\theta
d\theta}{\epsilon+b(\theta)\theta^2}}.$$ We now note that for
$\epsilon>0$
\begin{eqnarray*}-\frac{I(\epsilon)}{\epsilon}&=&\int_{0}^{\pi}{\frac{a(\theta)}{b(\theta)}\frac{\theta
d\theta}{\epsilon+b(\theta)\theta^2}} +
\int_{-\pi}^{0}{\frac{a(\theta)}{b(\theta)}\frac{\theta
d\theta}{\epsilon+b(\theta)\theta^2}}
\\&=&\int_{0}^{\pi}{\frac{a(\theta)}{b(\theta)}\frac{\theta
d\theta}{\epsilon+b(\theta)\theta^2}}
+\int_{\pi}^{0}{\frac{a(-\theta)}{b(-\theta)}\frac{\theta
d\theta}{\epsilon+b(-\theta)\theta^2}}
\\&=&\int_{0}^{\pi}{\theta\Big[\frac{a(\theta)}{b(\theta)}
\frac{1}{\epsilon+b(\theta)\theta^2}-
\frac{a(-\theta)}{b(-\theta)}
\frac{1}{\epsilon+b(-\theta)\theta^2}\Big] d\theta}.
\end{eqnarray*}
Thus if we define
\begin{equation}\label{defoj1}
J(\epsilon)= \int_{0}^{\pi}{\theta\Big[\frac{a(\theta)}{b(\theta)}
\frac{1}{\epsilon+b(\theta)\theta^2}-
\frac{a(-\theta)}{b(-\theta)}
\frac{1}{\epsilon+b(-\theta)\theta^2}\Big] d\theta},
\end{equation}
we have $I(\epsilon)=-\epsilon J(\epsilon)$ for $\epsilon>0$. The
important point now is that $J(\epsilon)$ makes sense for
$\epsilon=0$, indeed
\begin{equation}\label{j10}
J(0)=
\int_{0}^{\pi}{\frac{1}{\theta}\Big[\frac{a(\theta)}{(b(\theta))^2}
- \frac{a(-\theta)}{(b(-\theta))^2}\Big] d\theta},
\end{equation}
so that the function in the square brackets has a zero at
$\theta=0$, which implies that the integrand is continuous on
$[0,\pi]$, hence the integral is well-defined. Thus, if we can
ensure that
\begin{equation}\label{lim1}
\lim_{\epsilon\rightarrow 0+}{J(\epsilon)}=J(0)
\end{equation}
then we will have
\begin{equation}\label{lim2}
I'(0)=\lim_{\epsilon\rightarrow
0+}{\frac{I(\epsilon)}{\epsilon}}=-J(0).
\end{equation}
To prove (\ref{lim1}) we use Lebesgue's dominated convergence
theorem. The fact that the integrand of (\ref{defoj1}) converges
pointwise to the integrand of (\ref{j10}) as $\epsilon\rightarrow
0+$ is immediate. Let us show that the integrand of (\ref{defoj1})
is uniformly bounded:
\begin{eqnarray*} &&\Big|
\theta\Big[\frac{a(\theta)}{b(\theta)}
\frac{1}{\epsilon+b(\theta)\theta^2}-
\frac{a(-\theta)}{b(-\theta)}
\frac{1}{\epsilon+b(-\theta)\theta^2}\Big] \Big|\\ &\leq& \Big|
\frac{\theta a(\theta)}{b(\theta)}\Big[
\frac{1}{\epsilon+b(\theta)\theta^2}-\frac{1}{\epsilon+b(-\theta)\theta^2}
\Big]\Big|+\Big| \frac{\theta}{\epsilon+b(-\theta)\theta^2} \Big[
\frac{ a(\theta)}{b(\theta)}-\frac{
a(-\theta)}{b(-\theta)}\Big]\Big| \\&=& \Big| \frac{
a(\theta)}{b(\theta)} \frac{\theta^3(b(-\theta)-b(\theta))}
{(\epsilon+b(\theta)\theta^2)(\epsilon+b(-\theta)\theta^2)}\Big|+
\Big| \frac{\theta}{\epsilon+b(-\theta)\theta^2} \Big[ \frac{
a(\theta)}{b(\theta)}-\frac{ a(-\theta)}{b(-\theta)}\Big]\Big| \\
&\leq& \Big| \frac{ a(\theta)}{b(\theta)}
\frac{\theta^3(b(-\theta)-b(\theta))}
{(b(\theta)\theta^2)(b(-\theta)\theta^2)}\Big|+ \Big|
\frac{\theta}{b(-\theta)\theta^2} \Big[ \frac{
a(\theta)}{b(\theta)}-\frac{ a(-\theta)}{b(-\theta)}\Big]\Big|\\
&=& \Big| \frac{
a(\theta)}{(b(\theta))^2b(-\theta)}\frac{b(-\theta)-b(\theta)}{\theta}\Big|
+ \Big| \frac{1}{(b(-\theta))^2
b(\theta)}\frac{a(\theta)b(-\theta)-
b(\theta)a(-\theta)}{\theta}\Big|.
\end{eqnarray*}
Both functions on the right-hand side of the inequality are
continuous on $[0,\pi]$ since the numerators vanish at $\theta=0$,
so we have the uniform bound, and we have (\ref{lim2}), and thus
for small $\epsilon>0$ we have
$$sign(\chi)=sign(I(\epsilon))=sign(I'(0))=-sign(J(0)).$$
To complete the proof of theorem \ref{gent} we only need to show
that $J(0)=-\mu$ as defined by (\ref{defmu}). And indeed
\begin{eqnarray*}&-&\int_{-\pi+\gamma}^{\pi-\gamma}{\frac{h(\theta)}{w(\theta)}\Big(\frac{1}{P(\theta)
}\Big)'d\theta}=\int_{-\pi+\gamma}^{\pi-\gamma}{\frac{h(\theta)
P'(\theta)}{w(\theta)(P(\theta))^2}d\theta}=
\int_{\gamma}^{2\pi-\gamma}{\frac{h(\theta)
P'(\theta)}{w(\theta)(P(\theta))^2}d\theta}\\&=&
\int_{\gamma}^{2\pi-\gamma}{\frac{A(\theta)}{(B(\theta))^2}d\theta}=
\int_{-\pi+\gamma}^{\pi-\gamma}{\frac{1}{\theta}\frac{a(\theta)d\theta}{(b(\theta))^2}}
=\int_{0}^{\pi-\gamma}{\frac{1}{\theta}\Big[\frac{a(\theta)}{(b(\theta))^2}
- \frac{a(-\theta)}{(b(-\theta))^2}\Big] d\theta}. \end{eqnarray*}
Going to the limit $\gamma\rightarrow 0+$, we obtain $-\mu$ on the
left-hand side and (since the integrand on the right-hand side is
a continuous function) $J(0)$ on the right hand side (note that in
particular this proves that the improper integral defining $\mu$
exists).

\end{document}